\documentclass[letterpaper]{natureprintstyle}

\usepackage{graphicx}
\usepackage{amsmath, amssymb}
\usepackage{multirow}
\usepackage[varg]{txfonts}
\usepackage{ulem}
\usepackage{fancyhdr}

\newcommand{\ie}{i.e.}
\newcommand{\eg}{e.g.}

\newcommand{\natbox}[2]{
\begin{table*}[t!]
\colorbox{boxbl1}{
\begin{tabular}{p{0.96\textwidth}}
   {\sffamily\textbf{#1.}}
\end{tabular}
}

\vspace{-0.1cm}

\colorbox{boxbl2}{
\begin{tabular}{p{0.96\textwidth}}
#2
\end{tabular}
}
\end{table*}
}

\newcommand{\addresses}[1]{
\thispagestyle{fancy}
\lfoot{\parbox{\textwidth}{
\vspace{0.3cm}
   \rule{\textwidth}{0.2pt}
\hspace{-0.2cm}
\textsf{\scalefont{0.95} #1}
\vspace{-0.2cm}
\begin{center}{\scalefont{0.87} \thepage}\end{center}}}
\cfoot{}
}

\newcommand{\vect}[1]{{\bf #1}}

\newcommand{\new}[1]{}

\title{Disordered quantum gases under control}

\author{Laurent Sanchez-Palencia$^{1}$ and Maciej Lewenstein$^{2}$}

\date{When attempting to understand the role of disorder in condensed-matter physics, one faces severe experimental and theoretical difficulties and many questions are still open. Two of the most challenging ones, which have been debated for decades, concern the effect of disorder on superconductivity and quantum magnetism. Recent progress in ultracold atomic gases paves the way towards realization of versatile quantum simulators which will be useful to solve these questions. In addition, ultracold gases offer original situations and viewpoints, which open new perspectives to the field of disordered systems.}

\begin{document}

\maketitle
\addresses{
$^1$Laboratoire Charles Fabry de l'Institut d'Optique, CNRS and Univ. Paris-Sud, Campus Polytechnique, RD 128, F-91127 Palaiseau cedex, France;
$^2$ICREA - Instituci\'o Catalana de Recerca i Estudis Avan\c cats and ICFO - Institut de Ci\`encies Fot\`oniques, Parc Mediterrani de la Tecnologia, E-08860 Castelldefels (Barcelona)}

\noindent
\lettrine{P}hase coherence and interference effects underlie many basic phenomena in mesoscopic physics, for instance electronic conduction\cite{ashcroft1976}, magnetism\cite{auerbach1994}, superfluidity and superconductivity\cite{lifshitz1980}, or the propagation of light and sound waves in inhomogeneous media\cite{akkermans2006}. Both also play central roles in high-precision devices such as interferometers, accelerometers and gyroscopes. In this respect, an important issue concerns the effects of disorder, \ie\ of small random impurities, which cannot be completely avoided in real-life systems. A priori, one may expect that weak disorder slightly affects most physical systems and that averaging over the disorder smoothens possible effects. One may also expect that, in quantum systems, the spatial extension of wavefunctions leads to even weaker effects, via a kind of self-averaging. In fact, these naive ideas turn out to be wrong. Disorder often leads to subtle situations in which strong effects survive averaging over the disorder\cite{akkermans2006}, in particular in the quantum world. One of the most celebrated examples is Anderson localization\cite{anderson1958} (AL). It is now understood that AL results from interference of the many paths associated to coherent multiple scattering from random impurities, yielding wavefunctions with exponentially decaying tails and absence of diffusion\cite{lee1985}. This strongly contrasts with the Drude-Boltzmann theory of classical transport, which predicts that incoherent scattering induces diffusion\cite{ashcroft1976}.

Anderson localization was first introduced for non-interacting quantum particles to explain the absence of electronic conduction in certain dirty solids\cite{anderson1958}, but remained elusive for matterwaves. It was realized later that it is actually ubiquitous in wave physics\cite{akkermans2006}, paving the way for the first observations of AL, using classical waves, \eg\ light in diffusive media\cite{wiersma1997,storzer2006} and photonic crystals\cite{schwartz2007,lahini2008}, microwaves\cite{chabanov2000} and sound waves\cite{hu2008}. In condensed-matter physics, AL is now considered a fundamental phenomenon underlying certain metal-insulator transitions, but complete theory of disordered solids should incorporate Coulomb interaction, the underlying crystal structure, interaction with phonons, and magnetic effects. Unfortunately, understanding the physics of even the simplest models including all ingredients poses severe difficulties and many issues are still unsolved or even controversial. The most challenging ones concern the interplay of disorder and inter-particle interactions, and spin-exchange couplings.

Surprisingly enough, atomic physics offers new approaches to these issues. The field of ultracold atoms has been developing rapidly in the past decades, making it possible to produce, probe and manipulate Bose\cite{dalfovo1999,ketterle1999} and Fermi\cite{giorgini2008,ketterle2008} gases with unprecedented versatility, tunability and measurement possibilities (Box~1). Control in these systems is now such that ultracold atoms can realize quantum simulators\cite{feynman1982,cirac2004}, \ie\ platforms to investigate various fundamental models\cite{jaksch1998,bloch2005,lewenstein2007,bloch2008}. Landmark results have already been obtained, \eg\ observation of Mott insulators\cite{greiner2002,jordens2008,schneider2008}, Tonks-Girardeau\cite{paredes2004,kinoshita2004}, Berezinskii-Kosterlitz-Thouless\cite{hadzibabic2006} physics, and magnetic-like exchange\cite{anderlini2007,trotzky2008}.
Investigation of Bose-Einstein condensates (BECs) in disordered potentials\cite{clement2006,fallani2008} has also emerged in a quest for direct signatures of AL of matter-waves. Joint theoretical\cite{damski2003,roth2003,lsp2007,lsp2008} and experimental efforts\cite{lye2005,clement2005,fort2005,schulte2005,clement2008,chen2008,edwards2008} made it possible and two groups succeeded recently in observing one-dimensional AL\cite{billy2008,roati2008}.

\textit{Prima facie}, the discovery of this `Holy Grail' might mean the end of a quest. On the contrary, it is just a beginning as the two experiments of refs.~\citen{billy2008,roati2008} open unprecedented paths to pursue many outstanding challenges in the field of disordered systems. Direct extensions include studies of metal-insulator transitions in dimensions larger than one, and of the effect of weak interactions on localization, for which many questions are debated. For stronger interactions, single-particle localization is usually destroyed, but new concepts such as many-body Anderson localization\cite{basko2006,bilas2006,lugan2007bogo} and Bose glass\cite{giamarchi1988,fisher1989,scalettar1991,krauth1991} provide original paradigms, which renew our understanding of these issues. Experiments on ultracold atoms with controlled disorder and controlled interactions can also be extended to other systems where disorder plays important roles. For instance, combining spin exchange implementation\cite{anderlini2007,trotzky2008} and disorder opens the route towards random field-induced order\cite{minchau1985,wehr2006,niederberger2008} and spin glasses\cite{mezard1987,newman2003,sanpera2004,ahufinger2005}. These few examples illustrate all the promises of an emerging field, \ie\ quantum gases in controlled disorder. In this paper, we review theoretical and experimental progress in this area and discuss perspectives that are now within our grasp.

\section*{The nature of Anderson localization \\}
Localization, as introduced by P.W.~Anderson in 1958, is strictly speaking a single-particle effect\cite{anderson1958}.
Consider the wavefunction $\psi(\vect{r})$ of a free particle of mass $m$ and energy $E$, in a $d$-dimensional quenched disordered potential $V(\vect{r})$, which is solution of
the Schr\"odinger equation
\begin{equation}
E\psi(\vect{r}) = -\frac{\hbar^2\nabla^2}{2m}\psi(\vect{r}) + V(\vect{r})\psi(\vect{r}).
\label{eq:schrodinger}
\end{equation}
While in free space, $\psi(\vect{r})$ is an extended plane wave, it can be shown rigorously\cite{mott1961,borland1963} that, in the presence of disorder, any solution with arbitrary $E$ is exponentially localized in 1D, \ie\
$\ln(|\psi(z)|) \sim |z|/L_{\textrm{loc}}$, with localization length $L_{\textrm{loc}}(E) \propto l_{\textrm{B}}$, where $l_{\textrm{B}}$ is the transport (Boltzmann) mean-free path. Eventhough $L_{\textrm{loc}}$ often increases with $E$, it is striking that interference effects of multiply scattered waves is strong enough to profoundly affect $\psi(z)$, even for very high energies. In 2D, the situation is similar\cite{abrahams1979}, but interference effects are weaker, and $L_{\textrm{loc}} \propto l_{\textrm{B}}\exp(\pi k l_{\textrm{B}}/2)$ where $k = \sqrt{2mE}/\hbar$ would be the particle wavevector in free space. Hence $L_{\textrm{loc}}$ increases exponentially for $k > 1/l_{\textrm{B}}$, inducing a crossover from extended to localized states in finite-size systems. The situation differs dramatically in 3D where a proper phase transition occurs at the so-called mobility edge $k_{\textrm{mob}}$: While low-energy states with $k < k_{\textrm{mob}}$ are exponentially localized, those with $k > k_{\textrm{mob}}$ are extended. The exact features of the mobility edge are unknown, but approximately captured by the Ioffe-Regel criterion\cite{ioffe1960}, which basically states that localization requires the coherence volume contain several scattering processes. In other words, coherence must survive on longer distances than the memory of the initial particle direction, thus yielding $k_{\textrm{mob}} \sim 1/l_{\textrm{B}}$.

\section*{Direct observation of Anderson localization of matter-waves \\}
Observing AL of matter-waves requires several challenging conditions. First, one must use weak-enough disorder so that interference effects at the origin of AL dominate over classical trapping in potential minima. Second, one must eliminate all perturbations such as time-dependent fluctuations of the medium, or inter-particle interactions. Finally, one must demonstrate exponential localization, not only suppression of transport as it can also arise from classical trapping. While these conditions are very demanding in condensed-matter physics, they can be accurately fulfilled with ultracold atoms, using i)~controlled disorder, ii)~negligible interactions, iii)~strong isolation from the environment, and iv)~direct imaging of atomic density profiles. This way, direct signatures of AL of non-interacting matter-waves were reported in refs.~\citen{billy2008,roati2008}. As we shall see, these two experiments are complementary rather than similar because they significantly differ as regards both observation scheme and class of disorder.

\begin{figure}[!t]
\begin{center}
\vspace{6pt}
\includegraphics[width = 0.48\textwidth]{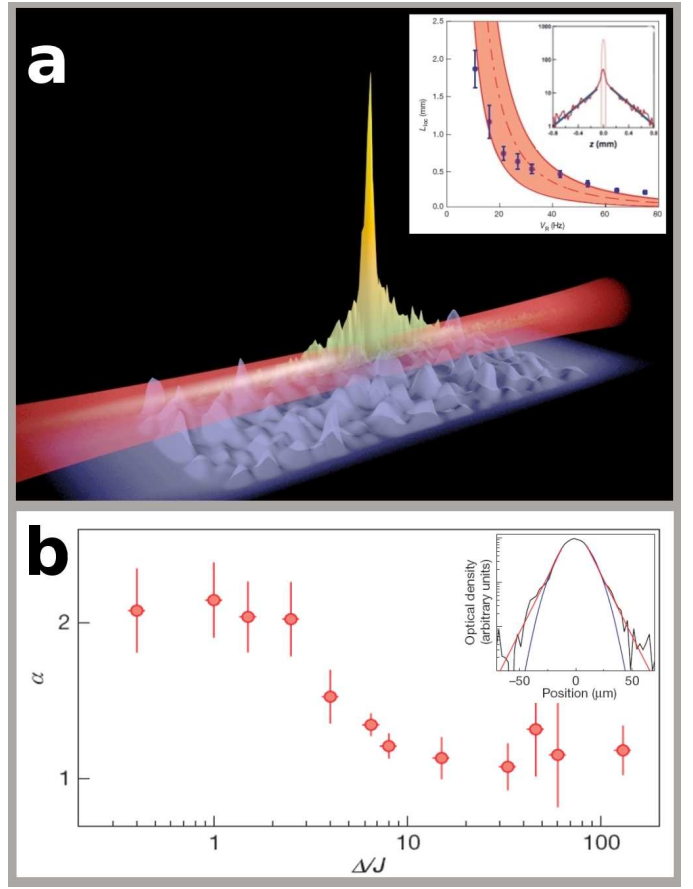}
\end{center}
\caption{
\textbf{Experimental observation of Anderson localization of matterwaves with Bose-Einstein condensates.}
\textbf{a) Experiments of Institut d'Optique} (coutesy of V.~Josse and P.~Bouyer): An interacting BEC expands in a tight 1D guide (in red) in the presence of a speckle potential (in blue). The expansion stops in less than 500ms and the density profile of the condensate is directly imaged (shown in orange-green; from the data of ref.~\citen{billy2008}). The column density, plotted in semi-logarithmic scale in the inset, shows a clear exponential decay characteristic of Anderson localization. The localization length $L_{\textrm{loc}}$, extracted by fitting an exponential $\exp(-2|z|/L_{\textrm{loc}})$ to the experimental profiles\cite{billy2008}, shows a good agreement with theoretical calculations\cite{lsp2007,lsp2008}.
\textbf{b) Experiments of LENS} (adapted from ref.~\citen{roati2008} with permission of the authors): A non-interacting BEC is created in a combination of a harmonic trap and a 1D bichromatic lattice. The plot shows the exponent $\alpha$ of a fit of a function $\exp(-|(x-x_0)/l|^\alpha)$ to the tails of the condensate at equilibrium in the combined potential, versus the ratio of the disorder strength ($\Delta$) to the site-to-site tunnelling rate ($J$). The onset of localization corresponds to the crossover to $\alpha \rightarrow 1$ for $\Delta/J > 9$. The inset shows a plot of the density profile of the condensate together with the fit for $\Delta/J = 15$.
}
\label{fig:AL}
\end{figure}

In ref.~\citen{billy2008}, a weakly interacting BEC is created in a trap, which is abruptly switched off at time $t = 0$. Then, the condensate expands in a guide and in the presence of disorder (Fig.~\ref{fig:AL}a), created with optical speckle (Box~2). This physics is captured by the Gross-Pitaevskii equation
\begin{equation}
\i\hbar \frac{\partial\psi}{\partial t} = -\frac{\hbar^2\nabla^2}{2m}\psi + V(\vect{r})\psi + g |\psi|^2\psi,
\label{eq:gpe}
\end{equation}
which corresponds to Box~1~Hamiltonian~(1) in meanfield regime. The dynamics of the BEC can be understood in a two-stage scheme\cite{lsp2007,lsp2008}.
First, it is dominated by interactions and the BEC expands, creating a coherent wavefunction with a stationary momentum distribution, $D(k) \propto 1-(k\xi)^2$, where $\xi = \hbar/\sqrt{4m\mu}$ is the initial healing length, which measures the initial interaction strength\cite{dalfovo1999}.
Second, once the expansion has strongly lowered the atomic density $\vert\psi(z)\vert^2$, the interaction term vanishes and we are left with a superposition of (almost) non-interacting waves $\psi_k$, the population of each is $D(k)$. Then each $\psi_k$ eventually localizes by interacting with the disordered potential, so that $\ln(\vert \psi_k (z)\vert) \sim \vert z \vert/L_{\textrm{loc}}(k)$, and the total BEC density reduces to\cite{lsp2007,lsp2008}
$n_{\textrm{BEC}}(z) \simeq \int dk\ D(k)\langle\vert \psi_k(z)\vert^2\rangle$.
Direct imaging of the localized matter-wave reveals exponentially decaying tails\cite{billy2008}, with a localization length equal to that of a non-interacting particle with $k = 1/\xi$ (Inset of Fig.~\ref{fig:AL}a). Hence, this experiment corresponds to a `transport scheme', which probes AL of non-interacting particles with a wavevector controlled by the initial interaction, via $\xi$.

In contrast, the experiment of ref.~\citen{roati2008} uses to a `static scheme'. The interactions are switched off already in the trap via Feshbach resonances, so that the gas is created in a superposition of a few (typically 1 to 3) low-energy, single-particle eigenstates. They are subsequently imaged \textit{in situ}, revealing exponentially decaying tails (Fig.~\ref{fig:AL}b). It is worth noting that ref.~\citen{roati2008} uses a 1D quasi-periodic, incommensurate lattice (Box~2), thus realizing the celebrated Aubry-Andr\'e model\cite{harper1955,aubry1980}
\begin{equation}
\hat{H} =
- \sum_{\langle j,l \rangle}
     J \left( \hat{a}^\dagger_{j}\hat{a}_{l} + \textrm{h.c.}\right)
+ \sum_{j}
     \Delta \cos (2\pi\beta j + \phi)\ \hat{a}^\dagger_{j}\hat{a}_{j}
\label{eq:AA}
\end{equation}
\ie\ Box~1~Eq.~(2) with $U = 0$, $V_j = \Delta \cos(2\pi \beta j + \phi)$, and $\beta$ an irrational number. Differently from the case of truly disordered potentials, there is a metal-insulator transition (mobility edge) in 1D, which is theoretically expected at $\Delta/J = 2$.

These works open new horizons to further deepen our knowledge of AL in various directions. In 1D, although all states are localized, subtle effects arise in correlated disorder, for instance in speckle potentials\cite{lsp2007}. To lowest order in the disorder amplitude,  $V_\textrm{R} = \sqrt{\langle V(z)^2\rangle}$, the Lyapunov exponent, $\gamma (k) = 1/L_{\textrm{loc}}(k)$, can be calculated analytically\cite{lifshits1988} and one finds $\gamma (k) \propto \langle V(2k)V(-2k)\rangle/k^2$, enlightening the role of coherent second-order back-scattering, $+k \rightarrow -k \rightarrow +k$, in the localization process. Since the power spectrum of speckle potentials has a cut-off $k_\textrm{C}$, such that $C_2(2k) = \langle V(2k) V(-2k) \rangle = 0$ for $k > k_\textrm{C}$ (Box~2), one finds an abrupt change (effective mobility edge) in the $k$-dependence of $\gamma$ for weak disorder\cite{lugan2009,gurevich2009}: While $\gamma (k) \sim V_\textrm{R}^2$ for $k < k_\textrm{C}$, higher-order scattering processes dominate for $k > k_\textrm{C}$ and $\gamma (k) \sim V_\textrm{R}^4$.

In dimensions higher than one, the self-consistent theory of localization\cite{vollhardt1980} allows one to calculate $L_{\textrm{loc}}$ and exhibits a mobility edge in 3D. It is however known that it is not fully exact. Therefore, a major challenge for disordered, ultracold atoms is to extend the works of refs.~\citen{billy2008,roati2008} to two\cite{kuhn2005,shapiro2007} and three\cite{skipetrov2008} dimensions. Definitely, observing the 3D mobility edge would be a landmark result, which may stimulate further theoretical developments and drive new approaches by providing precise measurements of the mobility edge $k_{\textrm{mob}}$
and the corresponding critical exponents, which are unknown.

\natbox{Box~1~$\vert$~Ultracold quantum gases}{
\vspace{0.2cm}
{\sffamily\bfseries Creating and manipulating ultracold gases} \\
Ultracold quantum gases are dilute atomic systems that are cooled down to temperatures of the order of a few tens of nano-Kelvins and confined in immaterial traps using combinations of magnetic and optic fields\cite{ketterle1999,ketterle2008}. Owing to strong dilution, the prominent inter-particle interactions are two-body interactions while many-body interactions can often be ignored. At ultra-low temperatures, s-wave scattering dominates and the interaction is accurately described by a contact pseudo-potential\cite{dalfovo1999,giorgini2008}. In the general case of mixtures of atoms in different species (or different internal states), the physics is thus governed by the Hamiltonian
\begin{equation}
~~~~~~~~~~~~~~~ \hat{H} =
\sum_{\sigma} \int d\vect{r}\ \hat{\Psi}_{\sigma}^\dagger (\vect{r})
				\left[ -\frac{\hbar^2\nabla^2}{2m_{\sigma}} + V_{\sigma}(\vect{r}) \right]
			      \hat{\Psi}_{\sigma} (\vect{r}) \ + \
\sum_{\sigma,\sigma'} \frac{g_{\sigma,\sigma'}}{2} \int d\vect{r}\
			      \hat{\Psi}_{\sigma}^\dagger (\vect{r})
			      \hat{\Psi}_{\sigma'}^\dagger (\vect{r})
			      \hat{\Psi}_{\sigma'} (\vect{r})
			      \hat{\Psi}_{\sigma} (\vect{r})
~~~~~~~~~~~~~(1)
\nonumber
\end{equation}
where $\hat{\Psi}_{\sigma}$
and $m_\sigma$ are the field operator and the mass of an atom of species $\sigma$. The first integral in Box~1~Eq.~(1) represents the single-particle Hamiltonian where the potential $V_\sigma (\vect{r})$ is controlled by the configuration of the magnetic and/or optic fields (Box~1~Fig~1a). In most cases, it is nearly a harmonic trap\cite{ketterle1999,ketterle2008}
($V_{\sigma} (x,y,z) = \sum_{\zeta\in\{x,y,z\}} m_{\sigma}\omega_{\sigma,\zeta}^2 \zeta^2/2$), the anisotropy of which can be adjusted in experiments. For instance, making it strongly anisotropic offers the possibility to produce one-\cite{dettmer2001,richard2003} or two-\cite{hadzibabic2006} dimensional gases. Another useful possibility is to create a guide for the atoms using a strongly focused laser beam\cite{guerin2006}. The second integral in Box~1~Eq.~(1) represents the interaction operator where $g_{\sigma,\sigma'}$ is the coupling constant for interacting atoms of same or different species ($g_{\sigma,\sigma'}>0$ and $g_{\sigma,\sigma'}<0$ correspond to repulsive and attractive interactions, respectively). Interestingly, the value and the sign of $g_{\sigma,\sigma'}$ can be controlled in quantum gases using Feshbach resonances\cite{bloch2008}. \\ \\

{\sffamily\bfseries Optical lattices} \\
Considering different limits of Hamiltonian~(1) allows one to design various models initially introduced in the context of condensed-matter physics, but here in a controlled way. One important example is that of optical lattices, which are produced from the interference pattern of several laser beams\cite{bloch2005,lewenstein2007,bloch2008}. The matter-light interaction creates a periodic potential whose geometry and amplitude are determined by the laser configuration and intensity. Both can be controlled in experiments. For instance, using pairs of counter-propagating laser beams (Box~1~Fig~1b), the lattice potential reads
$V_{\sigma}^{\textrm{latt}} (x,y,z) = V_{\sigma}^0 \sum_{\zeta\in\{x,y,z\}} \cos(2k_\textrm{L}\zeta)$, where
$V_{\sigma}^0$ is the lattice depth and $k_{\textrm{L}}$ the laser wavevector. In deep lattices, the atoms get trapped at the periodically-arranged minima of the lattice potential (so-called lattice sites). They can jump from site to site via quantum tunnelling (with a rate $J$), and two atoms interact only in the same site (with an energy $U$). This physics is governed by the Hubbard Hamiltonian, \ie\ the discrete version of Hamiltonian~(1):
\begin{equation}
~~~~~~~~~~~~~~~~~~~~ \hat{H} =
- \sum_{\sigma, \langle j,l \rangle}
     J_{\sigma} \left( \hat{a}^\dagger_{\sigma, j}\hat{a}_{\sigma, l} + \textrm{h.c.}\right)
\ + \
\sum_{\sigma, j}
     V_{\sigma, j}\ \hat{a}^\dagger_{\sigma, j} \hat{a}_{\sigma, j}
\ + \
\frac{1}{2} \sum_{\sigma, \sigma', j}
     U_{\sigma, \sigma'}\ \hat{a}^\dagger_{\sigma, j} \hat{a}^\dagger_{\sigma', j} \hat{a}_{\sigma', j} \hat{a}_{\sigma, j}
 ~~~~~~~~~~~~~~~~~~(2)
\nonumber
\end{equation}
where the sum over $\langle j,l\rangle$ covers all sites $j$ and their nearest-neighbour sites $l$, and
$\hat{a}_{\sigma,j}$ is the annihilation operator of an atom in site $j$. Hence, ultracold atoms (bosons or fermions) in optical lattices mimic the Hubbard model, which is widely considered in condensed-matter physics, for instance to capture the essential physics of electrons in solids. However, in contrast to condensed-matter systems, Hamiltonian~(2) can be shown to be exact in the limit of deep lattices, low temperature and low interactions\cite{jaksch1998}. The parameters $J_{\sigma}$, $V_{\sigma,j}$ and $U_{\sigma,\sigma',j}$ in Box~1~Eq.~(2) can be calculated \textit{ab initio} from the potential
$V_{\sigma} (\vect{r}) \rightarrow V_{\sigma} (\vect{r}) + V_{\sigma}^\textrm{latt} (\vect{r})$ and the coupling constant $g_{\sigma,\sigma'}$ in Box~1~Eq.~(1) and are thus controllable in experiments.
\vspace{0.5cm}
\begin{center}
\vspace{6pt}
\includegraphics[width = 0.95\textwidth]{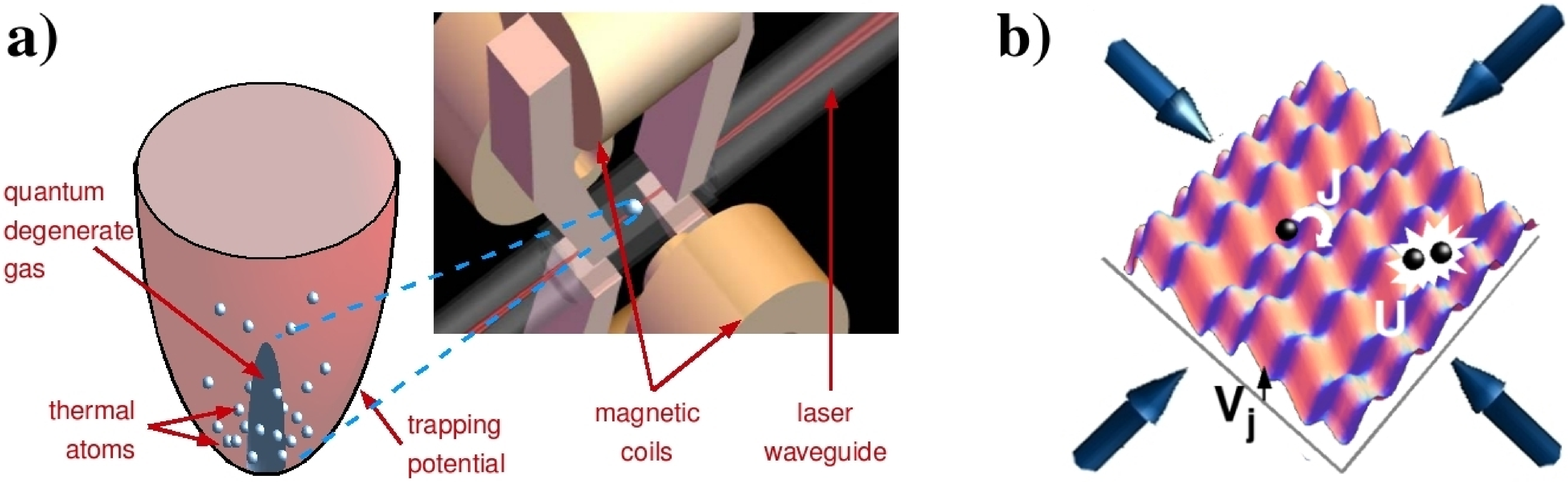}
\end{center}
\textbf{Box~1~Figure~1~$\vert$~Confining ultracold atoms in magnetic and optic traps.}
\textbf{a)~Harmonic trapping and laser waveguide} (coutesy of V.~Josse and P.~Bouyer).
Magnetic coils create a nearly harmonic trapping potential at the bottom of which a degenerate quantum gas, surrounded by a cloud of thermal atoms, is formed. A focused laser beam which creates an almost one-dimensional waveguide is also represented.
\textbf{b)~Optical lattice.} The interference pattern of pairs of counter-propagating laser beams form a periodic potential (represented here in two dimensions). The atoms are trapped in the lattice sites, but they can tunnel from site to site with a tunnelling rate $J$ and interact when placed in the same site with an energy $U$.}

\section*{Interactions versus Anderson localization \\}
Another outstanding challenge is to understand how interactions affect localization, a question that has proved puzzling from the earliest times of AL\cite{anderson1977}, and which is still debated. Common belief is that even weak interactions destoy localization. Different approaches however provide apparently contradicting answers in different transport schemes. For instance, recent numerical calculations\cite{pikovsky2008,palpacelli2008} suggest that for expanding BECs, repulsive interactions destroy AL beyond a given threshold. Conversely, other recent results\cite{kopidakis2008} predict that localization should persist even in the presence of interactions. Finally, in transmission experiments (which amount to throw a mono-kinetic wavepacket to a disordered region and measure transmission), perturbative calculations and numerical results indicate that repulsive interactions decrease the localization length before completely destroying localization\cite{paul2007}. Since a non-linear term is naturally present in BECs (see last term of Eq.~(\ref{eq:gpe})), and can be controlled via Feshbach resonances\cite{roati2008}, transport experiments with interacting condensates are particularly promising to address this question.

\begin{figure}[!t]
\begin{center}
\vspace{6pt}
\includegraphics[width = 0.48\textwidth]{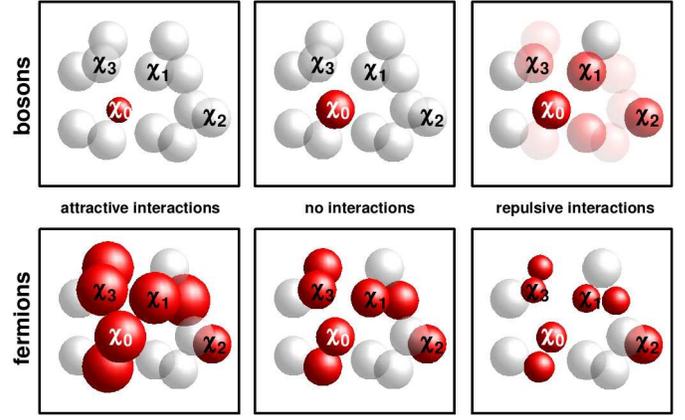}
\end{center}
\caption{
\textbf{Effect of interactions in disordered Bose and Fermi gases.}
The gas is described using the single-particle (non-interacting) states $\vert \chi_{\nu} \rangle$. In the presence of disorder, these states, which are localized and distributed around in a given volume, are represented by the spheres (in red when they are populated).
\textbf{Bose gas:}
For non-interacting bosons (top, central panel), the ground state, $\vert \chi_0 \rangle$, only is populated. Then, attractive interactions (top, left panel) tend to contract this state, thus favoring localization. Conversely, repulsive interactions (top, right panel) work against localization by populating more and more $\vert \chi_{\nu} \rangle$ states.
\textbf{Fermi gas:}
In the absence of interactions, a gas of $N$ fermions populates the $N$ lowest-energy $\vert \chi_{\nu} \rangle$ states (bottom, central panel). Then, each state tends to extend under the action of attractive interactions as for maximizing the overlap between different populated $\vert \chi_{\nu} \rangle$ states (bottom, left panel). Conversely, for repulsive interactions, they tend to minimize their mutual overlap,
then favoring localization (bottom, right panel).
}
\label{fig:interactions}
\end{figure}

A different approach to the interplay of interactions and localization consists in considering a Bose gas at equilibrium in a $d$-dimensional box of volume $\Omega$ in the presence of interactions and disorder (Fig.~\ref{fig:interactions}). For vanishing interactions and zero temperature, all bosons populate the single-particle ground state, $\vert \chi_0 \rangle$. Very weak attractive interactions are expected to favor localization by contracting the Bose gas, but also induce instabilities for moderate interactions, pretty much like for trapped BECs\cite{dalfovo1999}. Conversely, weak to moderate repulsive interactions do not affect much the stability, but work against localization, by populating an increasing number of single-particle states, $\vert \chi_{\nu} \rangle$. Weak interactions populate significantly only the lowest-energy states. Since they are strongly bound in rare, low-energy modulations of the potential, their mutual overlap is small. The gas then forms a Fock state,
$\vert \Psi \rangle \propto \prod_{\nu} (b_\nu^\dagger)^{N_\nu} \vert 0\rangle$, where 
$b_\nu^\dagger$ is the creation operator in state $\vert \chi_{\nu} \rangle$. The population $N_\nu$ of each is determined by the competition between single-particle energy $\epsilon_\nu$ and interaction within each state $\vert \chi_{\nu} \rangle$. This results in the characteristic equation of state\cite{lugan2007lifshits},
$Ng = \int^\mu d\epsilon\ \mathcal{D}_\Omega (\epsilon) (\mu-\epsilon) P(\epsilon)$,
where 
$\mathcal{D}_\Omega (\epsilon)$ is the density of states and
$P_{\nu} = 1/\int d \vect{r}\ \vert \chi_{\nu} (\vect{r}) \vert^4$ is the participation volume of $\vert \chi_{\nu} \rangle$. This state is an insulator with finite compressibility, $\kappa = \partial N/\partial \mu$, and can thus be refered to as a Bose glass\cite{giamarchi1988,fisher1989}. It attains particularly interesting features in disordered potentials bounded below (\ie\ when $V(\vect{r}) \gtrsim V_\textrm{min}$ everywhere), for which Lifshits has shown\cite{lifshitz1964} that the relevant single-particle states are determined by large-scale modulations of the potential. Since they are exponentially far apart, the density of state is exponentially small, $\mathcal{D}_\Omega (\epsilon) \sim \exp [-c (\epsilon-V_\textrm{min})^{-d/2}]$. As one can see, the equation of state is determined by both the density of state $\mathcal{D}_\Omega (\epsilon)$ and the localization via $P(\epsilon)$  in the Lifshits tail, which leads us to name this state the Lifshits-Anderson glass\cite{lugan2007lifshits}. In the opposite limit of strong interactions, there are very many populated $\vert \chi_{\nu} \rangle$, which thus overlap, and the above description breaks down. In turn, the gas forms an extended, connected (quasi-)BEC of density $n(\vect{r}) = [\mu-V(\vect{r})]/g$, which is well described in meanfield approach\cite{lsp2006}. This state is a superfluid. Finally, the intermediate region interpolates between the Lifshits-Anderson glass and the BEC regime. Then, the Bose gas separates in fragmented, forming a compressible insulator (Bose glass). The characteristic features of the fragments can then be estimated in the meanfield framework\cite{falco2009}.

The above description is consistent with the idea that even weak interactions destroy single-particle localization. In order to gain further insight, it is worth noting that in interacting systems, the relevant states are not the single-particle eigenstates, but are of collective nature. For interacting BECs, they are Bogolyubov quasi-particles\cite{dalfovo1999}. One then finds that, although the ground state is extended, the Bogolyubov quasiparticles are localized\cite{bilas2006,lugan2007bogo,gurarie2008}. Their localization properties however differ from those of Schr\"odinger particles, owing to a strong screening of disorder\cite{lugan2007bogo}. In 1D, the Lyapunov exponent of a Bogolyubov quasiparticle reads $\Gamma (k) = [\mathcal{S}(k)]^2 \gamma (k)$, where $\gamma (k)$ is the single-particle Lyapunov exponent and $\mathcal{S}(k) = 2(k\xi)^2/(1+2(k\xi)^2)$ is the screening function. One thus finds that in the phonon regime ($k \ll 1/\xi$), the screening is strong and $\Gamma (k) \ll \gamma (k)$. Conversely, in the free-particle regime ($k \gg 1/\xi$), the screening vanishes and $\Gamma (k) \simeq \gamma (k)$. Hence, surprisingly, localization can survive in the presence of strong mean-field interactions. This poses new challenges to ultracold atoms: Not only direct observation of many-body AL, but also possible consequences on quantum coherence, sound-wave propagation or thermalization process.

\natbox{Box~2~$\vert$~Creating controlled disordered potentials}{
In atomic gases, disorder can be created in a controlled way. For instance, the so-called speckle potentials are formed as follows\cite{goodman2007}. A coherent laser beam is diffracted through a ground-glass plate and focused by a converging lens (Box~2~Fig.~1a). The ground-glass plate transmits the laser light without altering the intensity, but imprinting a random phase profile on the emerging light. Then, the complex electric field $\mathcal{E}(\vect{r})$ on the focal plane results from the coherent superposition of many waves with equally-distributed random phases, and is thus a Gaussian random process. In such a light field, atoms with a resonance slightly detuned with respect to the laser light experience a disordered potential $V(\vect{r})$ which, up to a shift introduced to ensure that the statistical average $\langle V \rangle$ of $V(\vect{r})$ vanishes, is proportional to the light intensity,
$V(\vect{r}) \propto \pm (\vert \mathcal{E}(\vect{r})\vert^2-\langle \vert \mathcal{E}\vert^2\rangle)$,
an example of which in shown in Box~2~Fig.~1b. Hence, the laws of optics allows us to precisely determine all statistical properties of speckle potentials. First, although the electric field $\mathcal{E}(\vect{r})$ is a complex Gaussian random process, the disordered potential $V(\vect{r})$ is not Gaussian itself, and its single-point probability distribution is a truncated, exponential decaying function,
$P\left(V(\vect{r})\right) = \textrm{e}^{-1}\vert V_\textrm{R}\vert^{-1}\exp(-V(\vect{r})/V_\textrm{R})\Theta\left(V(\vect{r})/V_\textrm{R}+1\right)$,
where $\sqrt{\langle V^2\rangle} = \vert V_\textrm{R}\vert$ is the disorder amplitude and $\Theta$ is the Heaviside function. Both modulus and sign of $V_\textrm{R}$ can be controlled experimentally\cite{clement2006}: The modulus is proportional to the incident laser intensity while the sign is determined by the detuning of the laser relative to the atomic resonance ($V_\textrm{R}$ is positive for `blue-detuned' laser light\cite{clement2006,clement2005,clement2008,billy2008}, and negative for `red-detuned' laser light\cite{lye2005,fort2005,chen2008}). Second, the two-point correlation function of the disordered potential,
$C_2(\vect{r}) = \langle V(\vect{r}) V(0)\rangle$, is determined by the overall shape of the ground-glass plate but not by the details of its asperities\cite{goodman2007}. It is thus also controllable experimentally\cite{clement2006}. There is however a fundamental constraint: Since speckle potentials result from interference between light waves of wavelength $\lambda_\textrm{L}$ coming from a finite-size aperture of angular width $2\alpha$ (Box~2~Fig.~1a) they do not contain Fourier components beyond a value $2k_\textrm{C}$, where $k_\textrm{C} = (2\pi/\lambda_\textrm{L})\sin(\alpha)$.
In other words, $C_2(2\vect{k}) = 0$ for $\vert \vect{k} \vert > k_\textrm{C}$.

\hspace{0.25cm}
Speckle potentials can be used directly to investigate the transport of matter-waves in disordered potentials\cite{lye2005,clement2005,fort2005,schulte2005}. They can also be superimposed to deep optical lattices\cite{white2009}. In the latter case, the physics is described by Box~1~Hamiltonian~(2) with $V_{\sigma,j}$ a random variable whose statistical properties are determined by those of the speckle potential. In particular, $V_{\sigma,j}$ is non-symmetric and correlated from site to site. Yet another possibility to create disorder in deep optical lattices is to superimpose a shallow optical lattice with an incommensurate period\cite{schulte2005,edwards2008,roati2008,fallani2007}. In this case, $V_{\sigma,j} = \Delta \cos(2\pi \beta j + \phi)$, where $\Delta$ and $\phi$ are determined by the amplitude and the phase of the second lattice and $\beta = k_2/k_1$ is the (irrational) ratio of the wavevectors of the two lattices. Although the quantity $V_{\sigma,j}$ is deterministic, it mimics disorder in finite-size systems\cite{damski2003,roth2003,roscilde2008,roux2008}. In contrast to speckle potentials, these bichromatic lattices form a pseudo-random potential, which is bounded ($\vert V_{\sigma,j}\vert \lesssim \Delta$) and symmetrically distributed.
\vspace{0.5cm}
\begin{center}
\vspace{6pt}
\includegraphics[width = 0.95\textwidth]{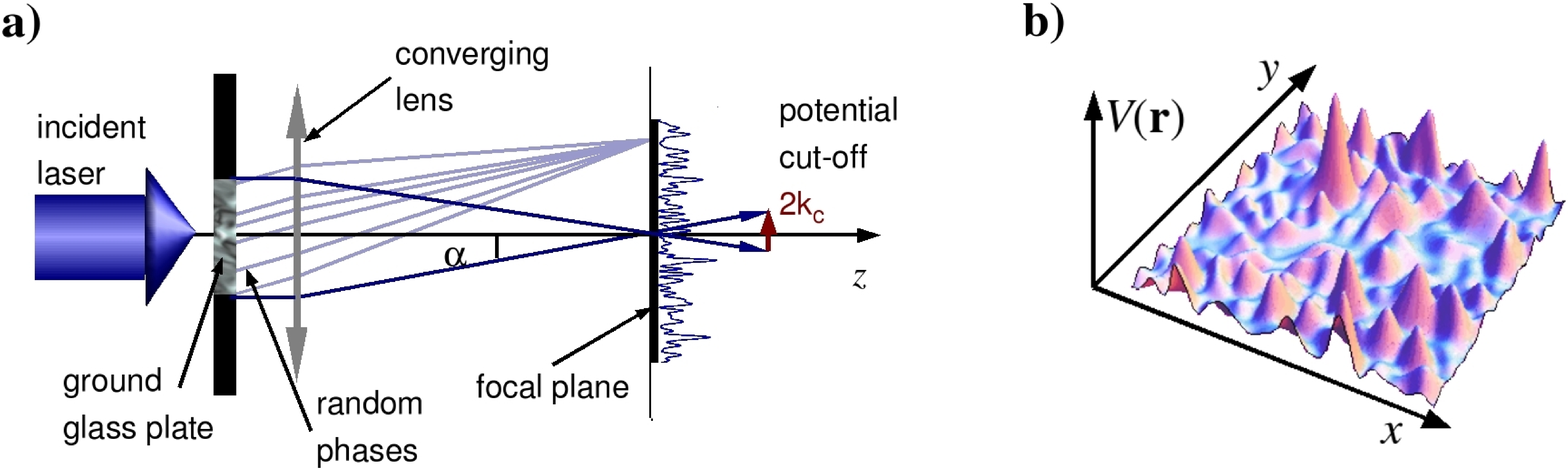}
\end{center}
\textbf{Box~2~Figure~1~$\vert$~Optical speckle potentials.}
\textbf{a)}~Optical configuration.
\textbf{b)}~Two-dimensional representation of a speckle potential.
}

\section*{Fermi systems and `dirty' superconductors \\}
Consider now a Fermi gas, and focus again on the ground state properties (Fig.~\ref{fig:interactions}). In the absence of interactions, the gas of $N$ fermions populates the $N$ lowest single-particle states. For low density, short-range interactions do not play a significant role as the populated states are spatially separated. However, for large-enough density, they do overlap. Then, for repulsive interactions, each populated state tends to contract to minimize its overlap with other populated states, thus favoring localization. Conversely, for attractive interactions, the populated states tend to extend to maximize their overlap, thus favoring delocalization. Hence strikingly, interactions have opposite consequences for fermions and bosons.

Perhaps even more fascinating is the possibility to study `dirty' Fermi liquids. Experiments with two-component Fermi gases (\eg\ $^6$Li or $^{40}$K), with interactions controlled by Feshbach resonances, have already significantly advanced our understanding of the so-called BEC-BCS crossover\cite{giorgini2008,ketterle2008}. On the attractive side of the resonance and for weak interactions, the Fermi superfluid is well described by the Bardeen-Schrieffer-Cooper (BCS) theory and formation of spatially extended Cooper pairs consisting of two fermions of opposite spins and momenta. On the repulsive side, pairs of fermions form bosonic molecules, which undergo Bose-Einstein condensation. Although disorder should not significantly affect pairing, BCS superfluidity and BEC superfluidity are expected to react differently to disorder\cite{orso2007,han2009}. The famous Anderson theorem\cite{anderson1959} indicates that disorder should not affect very much the BCS superfluid owing to the long-range and overlapping nature of the Cooper pairs. Conversely, disorder should seriously affect the molecular BEC, enhancing phase fluctuations.

\section*{Strongly-correlated gases in disordered lattices \\}
Strong interactions are also very important in various disordered systems, \eg\ superfluids in porous media or `dirty' superconductors. Metal-insulator transitions attain a particularly interesting, but not fully understood character in lattice systems. In this respect, the Bose-Hubbard model,
\begin{eqnarray}
\hat{H} & = &
- \sum_{\langle j,l \rangle}
     J \left( \hat{a}^\dagger_{j}\hat{a}_{l} + \textrm{h.c.}\right)
+ \sum_{j}
     V_{j}\ \hat{a}^\dagger_{j} \hat{a}_{j}
+ \frac{1}{2} \sum_{j}
     U\ \hat{a}^\dagger_{j} \hat{a}^\dagger_{j} \hat{a}_{j} \hat{a}_{j}
\label{eq:BoseHubbard}
\end{eqnarray}
is central in condensed-matter physics\cite{fisher1989,scalettar1991,krauth1991} for it forms a tractable model, which captures the elementary physics of strongly interacting systems. Hamiltonian~(\ref{eq:BoseHubbard}) describes bosons, in a lattice with inhomogeneous on-site energies $V_j$, which can tunnel between the sites, with rate $J$, and interact when placed in the same site, with interaction energy $U$. Interestingly, this model contains the most fundamental two phenomena underlying metal-insulator transitions. They correspond to the Anderson transition\cite{anderson1958,lee1985} in the absence of interactions ($U = 0$) as discussed above, and to the Mott transition\cite{mott1968} in the absence of disorder ($V = 0$). In systems dominated by repulsive interactions, density fluctuations, which are energetically costy, are suppressed, and a Mott insulator (MI) state, $\vert \Psi_{\textrm{MI}}\rangle \propto \prod_j (\hat{a}_j^\dagger)^n\vert 0 \rangle$, is formed. Then, the number of bosons per site, $n = [\mu/U+1]$, where $[.]$ represents the integer part, is determined and phase coherence between the lattice sites vanishes. MIs are insulating, incompressible, and gapped as the first excitation corresponds to transfer one atom from a given site to another, which costs the finite energy $U$. At the other extreme, when tunneling dominates, the bosons form a superfluid state,
$\vert \Psi_{\textrm{SF}}\rangle \propto \left(\sum_j \hat{a}_j^\dagger\right)^N\vert 0 \rangle$, with normal density fluctuations and perfect coherence between the lattice sites. This state is gapless and compressible.

In the presence of disorder, a glassy phase is formed, which interpolates between Lifshits-Anderson glass for weak interactions, to Bose glass for strong interactions\cite{fisher1989}. The latter can be represented as
$\vert \Psi_{\textrm{BG}}\rangle \propto \prod_j (\hat{a}_j^\dagger)^{n_j}\vert 0 \rangle$
with $n_j = [(\mu-V_j)/U+1]$. This phase is thus insulating but compressible and gapless since the ground state is quasi-degenerated, like in many other glassy systems\cite{fisher1989,scalettar1991,krauth1991}. With the possibility of realizing experimentally systems exactly described by Hubbard models (Box 1), ultracold atoms in optical lattices offer also here unprecedented opportunities to investigate this physics in detail, and to directly observe the Bose glass, which has not been possible in any system so far. Two experimental groups have made the first steps in this direction\cite{fallani2007,white2009}. The experiment of ref.~\citen{fallani2007} applied a bichromatic, incommensurate lattice to 1D Mott insulators. Increasing disorder, a broadening of Mott resonances was observed, suggesting vanishing of the gap and transition to an insulating state with a flat density of excitations. Intensive theoretical studies have been devoted to understand these results, using quantum Monte-Carlo\cite{roscilde2008} and Density Matrix Renormalization Group\cite{roux2008} techniques. The results of ref.~\citen{roscilde2008} suggest that, in the conditions of ref.~\citen{fallani2007}, one should expect a complex phase diagram with competing regions of gapped, incompressible band-insulator, and compressible Bose glass phases. Clearly, novel and more precise detection schemes are needed to characterize this kind of physics, such as direct measurements of compressibility\cite{scalettar1991} or condensate fraction in superfluid, or coexisting superfluid and MI phases. The latter has been approached experimentally in ref.~\citen{white2009}, where disorder-induced suppression of the condensate fraction in a lattice with super-imposed speckle was observed.

One can also investigate the corresponding Fermi counterparts with ultracold atoms. These systems are particularly interesting as they would mimic superconductors, even better than bosons. In this respect, an outstanding challenge is definitely to understand high-T$_\textrm{C}$ superconductors, and possibly important effects of disorder in these systems. Consider the two-component ($\sigma \in \{\uparrow,\downarrow\}$) Fermi-Hubbard Hamiltonian
\begin{eqnarray}
\hat{H} & = &
- \sum_{\sigma, \langle j,l \rangle}
     J_{\sigma} \left( \hat{a}^\dagger_{\sigma, j}\hat{a}_{\sigma, l} + \textrm{h.c.}\right)
+ \sum_{\sigma, j}
     V_{\sigma, j}\ \hat{a}^\dagger_{\sigma, j} \hat{a}_{\sigma, j}
\label{eq:FermiHubbard} \\
&& + \sum_{j}
     U\ \hat{a}^\dagger_{\uparrow, j} \hat{a}^\dagger_{\downarrow, j} \hat{a}_{\downarrow, j} \hat{a}_{\uparrow, j}
\nonumber
\end{eqnarray}
For weak interactions, we have a Fermi liquid similar to that discussed above. For strong interactions and low temperature, $T \lesssim U$, the Fermi gas enters a MI state, pretty much like for bosons, but with a single ($n = 1$) fermion per site (either $\uparrow$ or $\downarrow$). Evidence of vanishing double occupancy and incompressibility have been reported recently in Fermi MIs\cite{jordens2008,schneider2008}. Then, in the presence of disorder, various phases could be searched for, such as Fermi glasses. At even lower temperatures, spin exchange starts to play a role, and a transition from paramagnetic to antiferromagnetic insulator phases is predicted for $T_{\textrm{N}} \sim 4J^2/U$ in non-disordered systems\cite{georges2007}. Interestingly, the interplay of interactions and disorder might lead to appearance of novel `metallic' phases between the Fermi glass and the MI. Hence, dynamical mean-field theory\cite{byczuk2009} at half-filling predicts that disorder tends to stabilize paramagnetic and antiferromagnetic metallic phases for weak interactions. For strong interactions however, only the paramagnetic Anderson-Mott insulator (for strong disorder) and antiferromagnetic insulator (for weak disorder) phases survive.

\section*{Simulating disordered spin systems \\}
In condensed-matter physics, other important paradigm models where disorder induces subtle effects are
spin systems, described by the Hamiltonian
\begin{equation}
\hat{H} =
- \sum_{\langle j,l \rangle} \left(
     J_{j,l}^x \ \hat{S}^x_j \cdot \hat{S}^x_l
   + J_{j,l}^y \ \hat{S}^y_j \cdot \hat{S}^y_l
   + J_{j,l}^z \ \hat{S}^z_j \cdot \hat{S}^z_l
   \right)
\ - \ \sum_{j} \vect{h}_j \cdot \vect{\hat{S}}_j,
\label{eq:spins}
\end{equation}
with either random spin exchange, $J_{j,l}$, or random magnetic field, $\vect{h}_j$. Ultracold gases can also simulate this class of systems, although not as straightforwardly as for Hubbard models. Consider a two-component (Bose-Bose or Fermi-Bose) ultracold gas in an optical lattice, as described by Box~1~Hamiltonian~(2).
In the strongly-correlated regime, the couplings between the particles can be understood as exchange-mediated interactions between composite (bosonic or fermionic) particles.
One can then map Box~1~Hamiltonian~(2) onto Hamiltonian~(\ref{eq:spins}) with fictitious spins encoded in combinations of the annihilation and creation operators of the composite particles:
$\hat{S}_{j,x} = (\hat{A}_j+\hat{A}_j^\dagger)/2$,
$\hat{S}_{j,y} = (\hat{A}_j-\hat{A}_j^\dagger)/2i$,
and $\hat{S}_{j,z} = 1/2-\hat{A}^\dagger\hat{A}$, which indeed have commutation relations of spins\cite{auerbach1994}, and complicated but analytical functional dependence of $J_{j,l}$ and $\vect{h}_j$ on the parameters of Box~1~Hamiltonian~(2).
In the presence of disorder, these parameters are random\cite{wehr2006,ahufinger2005} and one can reach various limiting cases corresponding to Fermi glass, quantum spin glass and quantum percolation\cite{sanpera2004}.

\begin{figure}[!t]
\begin{center}
\vspace{6pt}
\includegraphics[width = 0.48\textwidth]{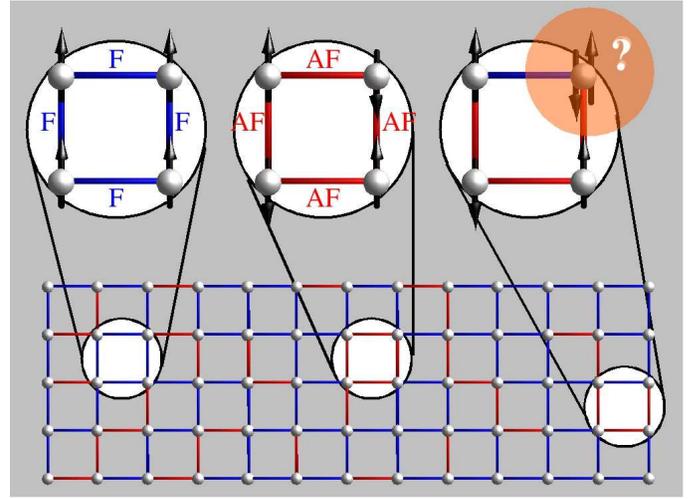}
\end{center}
\caption{
\textbf{The spin glass problem.}
An assembly of spins located at the nodes of a cubic lattice interacts according to Hamiltonian~(\ref{eq:spins}) where the exchange term $J_{j,l}$ only is randomly distributed, and can be either ferromagnetic (blue bonds) or anti-ferromagnetic (red bonds). The ground state of the system corresponds to the spin configuration that minimizes the total energy. The inherent complexity of spin glasses results from frustration which appears when the topography of ferromagnetic and anti-ferromagnetic bonds make impossible to fulfil the local constraints all together. In some plaquetes of four sites, local minimization is easy, for instance when all bonds are ferromagnetic (left disk) or anti-ferromagnetic (central disk). In some others, it leads to frustration, for instance for odd numbers of ferromagnetic and anti-ferromagnetic bonds (right disk). In the latter case, at least one spin is frustrated, that is its spin orientation is not unique. Hence, frustration is at the origin of a manifold of metastable states which corresponds to configurations with similar energies.
}
\label{fig:spinglass}
\end{figure}

Particularly promising is the possibility of simulating spin glasses\cite{sanpera2004} (Fig.~\ref{fig:spinglass}), for which only the exchange term, $J_{j,l}$, is randomly distributed. The phase diagram of (even classical) spin glasses, which is not known yet, is an outstanding challenge in condensed-matter physics. The nature of spin glasses is still debated and there exist competing theories: The Parisi replica symmetry breaking\cite{mezard1987} and the Nelson-Huse droplet model\cite{newman2003}. Ultracold atoms might contribute to the resolution of this issue, not only on the classical level but also on the quantum level since they offer original ways of performing quenched averages. Importantly with a view towards testing the replica theory, overlap between two spin configurations between two (or more) replicas can be measured directly by preparing a pair of 2D lattices with the same realization of disorder\cite{morrison2008}. Quenched averages for systems with binary disorder can also be simulated by replacing the classical disorder variables by quantum $1/2$-spins, and preparing them in a superposition state\cite{paredes2005}.

Yet another fascinating possibility is to simulate various random field-induced order (RFIO) phenomena in systems with continuous symmetry, such as BECs or XY-spin models with $U(1)$ symmetry, or Heisenberg models with $SU(2)$ symmetry\cite{wehr2006,niederberger2008}. A prototype model\cite{abanin2007}, is the 2D-XY version of Hamiltonian~(\ref{eq:spins}) with fixed exchange $J_{j,l}$ but random field $\vect{h}_j$. In the absence of disorder, symmetry leads to strong fluctuations, which suppress long-range order, according to the Mermin-Wagner-Hohenberg theorem. Disorder distributed in a symmetric way suppresses ordering even more. Surprisingly however, disorder that breaks symmetry might actually favor ordering. This model can be implemented within Bose-Bose mixtures\cite{wehr2006,niederberger2008} where random uniaxial $\vect{h}_j$ can be implemented using two internal states of the same atom, coupled via a random Raman field, $\hbar\Omega(\vect{r})\Psi_1(\vect{r})^\dagger\Psi_2(\vect{r}) + h.c.$
In order to break the continuous symmetry, one uses a Raman coupling with constant phase, but random strength. In lattice systems, RFIO shows up but is limited by finite-size effects, even in very large systems\cite{wehr2006}. In this respect, ultracold atoms offer an alternative and fruitful route. Indeed, RFIO turns out to be particularly efficient in two (or multi-) component BECs in meanfield regime, where the energy functional reads
$\Delta E \simeq d\vect{r}\ n [(\hbar^2/4m)(\nabla\theta)^2 + \hbar\Omega(\vect{r})\cos(\theta)]$,
with $n$ the atomic density and $\theta(\vect{r}) =  \theta_1(\vect{r}) - \theta_2(\vect{r})$, the phase difference between the two BECs. This is the continuous counterpart of the 2D-XY model. Then, RFIO manifests itself as a fixed $\theta(\vect{r}) = \pm\pi/2$, and thus allows to control the relative phase between the components\cite{niederberger2008}. This is a striking example where ultracold atoms can be used not only to simulate classic models, but also offer new and fruitful viewpoints to fundamental issues.

\section*{Further directions \\}
As the reader has probably noticed, we both are very enthusiastic about the future development of the field of disordered quantum gases, and probably would like that any interesting direction can be pursued. Limited size of the present review has not allowed us to discuss them all, but let us briefly mention another.

Two-component (Bose-Bose or Bose-Fermi) mixtures offer an alternative method to create disorder in optical lattices, namely by quenching one component in random sites, so as to form a background of randomly-distributed impurities\cite{paredes2005,gavish2005}. Theoretical analysis using Gutzwiller method confirms the appearance of incompressible MI and partially compressible Bose glass phases\cite{buonsante2007}. The idea of freezing the motion of the second species to form random impurities (\ie\ classical disorder) can be generalized to freezing of any quantum state\cite{horstmann2007}. In this case the system does not involve any classical disorder, but nevertheless localization occurs owing to quantum fluctuations in the frozen state of the second species.

One can even relax the freezing condition and consider say two bosonic species, one of which tunnels much slower than the other, forming a quasi-static disorder. In a large region of parameters (for repulsive inter-species forces), the ground state corresponds to full phase segregation. In practice it is marked by a large number of metastable states in which microscopic phase separation occurs, reminiscent of emulsions in immiscible fluids\cite{roscilde2007}. Such quantum emulsions are predicted to have very similar properties to the Bose glass phase, \ie\ compressibility and absence of superfluidity. Such quasi-static or even time-dependent disorder effects have been suggested to underlie the quite large shift of the SF-MI transition in Bose-Fermi\cite{ospelklaus2006,gunter2006} and Bose-Bose\cite{catani2008} mixtures. This issue was quite controversial and the most recent work suggests that, while indeed the fermions tend to localize the bosons for attractive
boson-fermion interactions, higher Bloch bands play a significant role\cite{luhmann2008,best2009,lutchyn2009}.


\section*{Acknowledgements \\}
This research was supported by the French Centre National de la Recherche Scientifique (CNRS), Agence Nationale de la Recherche (ANR), Triangle de la Physique and Institut Francilien de Recherche sur les Atomes Froids (IFRAF), the German Alexander von Humboldt foundation, the Spanish MEC grants (FIS 2005-04627, Conslider Ingenio 2010 "QOIT"), the European Union IP Programme SCALA and the European Science Foundation - MEC Euroquam Project FerMix.

\section*{Additional Information  \\}
The authors declare that they have no competing financial interests. Correspondence and requests should be addressed to L.S.P.~(lsp@institutoptique.fr) or M.L.~(maciej.lewenstein@icfo.es).

\end{document}